\documentstyle[twoside,fleqn,espcrc2,epsfig]{article}
\renewcommand{\aa}[1]{
 \unitlength#1in
 \begin{picture}(2,1)
 \multiput(0,0)(1,0){3}{\line(0,1){1}}
 \multiput(0,0)(0,1){2}{\line(1,0){2}}
 \end{picture}
}
\newcommand{\ab}[1]{
 \unitlength#1in
 \begin{picture}(1,2)
 \multiput(0,0)(1,0){2}{\line(0,1){2}}
 \multiput(0,0)(0,1){3}{\line(1,0){1}}
 \end{picture}
}
\newcommand{\aaaa}[1]{
 \unitlength#1in
 \begin{picture}(4,1)
 \multiput(0,0)(1,0){5}{\line(0,1){1}}
 \multiput(0,0)(0,1){2}{\line(1,0){4}}
 \end{picture}
}
\newcommand{\aaab}[1]{
 \unitlength#1in
 \begin{picture}(3,2)
 \multiput(0,0)(1,0){2}{\line(0,1){2}}
 \multiput(2,1)(1,0){2}{\line(0,1){1}}
 \multiput(0,0)(0,1){1}{\line(1,0){1}}
 \multiput(0,1)(0,1){2}{\line(1,0){3}}
 \end{picture}
}
\newcommand{\aabb}[1]{
 \unitlength#1in
 \begin{picture}(2,2)
 \multiput(0,0)(1,0){3}{\line(0,1){2}}
 \multiput(0,0)(0,1){3}{\line(1,0){2}}
 \end{picture}
}
\renewcommand{\slash}[1]{#1\!\!\!/}

\newcommand{\AmS}{{\protect\the\textfont2
  A\kern-.1667em\lower.5ex\hbox{M}\kern-.125emS}}

\title{Meson-meson scattering in the massive Schwinger model --- a 
       status report\thanks{Work supported by the Deutsche
       Forschungs\-gemein\-schaft.}
       \hfill \raisebox{4cm}[-4cm]{\normalsize Aachen, PITHA 97/39} 
      } 

\author{C.~Gutsfeld\address
                     {Institut f\"ur Theoretische Physik E,
                      RWTH Aachen, D-52056 Aachen, Germany
                     }$^{\rm,}$\address
                     {H\"ochstleistungsrechenzentrum HLRZ, 
                      c/o Forschungszentrum J\"ulich, 
                      D-52425 J\"ulich, Germany
                     },
        H.~A.~Kastrup$^{\rm a}$,
        K.~Stergios$^{\rm a, b,}$\thanks{Speaker at the conference.}
        and
        J.~Westphalen\address
                     {Deutsches Elektronen-Synchrotron DESY, 
                      Notkestra{\ss}e 85, D-22603 Hamburg, Germany
                     }
       }
       
\begin{document}

\begin{abstract}
We discuss the possibility of extracting phase shifts from finite 
volume energies for meson-meson scattering, where the mesons are 
fermion-antifermion bound states of the massive Schwinger model 
with SU(2) flavour symmetry. 
The existence of analytical strong coupling predictions for the 
mass spectrum and for the scattering phases makes it possible to 
test the reliability of numerical results. 
\end{abstract}

\maketitle

\section{INTRODUCTION}

After the successful applications of L\"uscher's method
\cite{luescher_91} to the elastic scattering of {\it elementary} 
scalar and fermionic particles (see, e.g., \cite{phi4_94,gn_96}), we 
have chosen the massive Schwinger model (QED$_2$) with 2 flavours 
to determine scattering phases in meson-meson scattering.
Similar to four-dimensional QCD, mesons occur because of confinement.
This two-dimensional model also exhibits charge screening and it 
possesses a nontrivial vacuum structure.
Moreover, there exist analytical predictions which enable us to test 
the numerical results.

\section{L\"USCHER'S RELATION IN d=2}
 
L\"uscher's relation connects elastic scattering phases in infinite
volume and two-particle energies in finite volumes. In 2 dimensions
it has the simple form
\begin{equation}
  2 \delta(k) = - k L \quad (\mbox{mod} \; 2\pi).
\label{eq:luescher_relation}
\end{equation}
Using the energy-momentum relation, $k$ can be calculated from the 
mass and the two-particle energies, which are accessible to Monte 
Carlo simulations.

\section{THE SCHWINGER MODEL}

The massive Schwinger model with 2 flavours has the following 
continuum Euclidean action:
\begin{eqnarray}
  S_{cont} &=& \int d^2x \{ \frac{1}{4}F_{\mu\nu}F^{\mu\nu} + 
\label{eq:S_cont}\\
           & & \quad \sum_{f=1}^{2} \bar{\psi}^{f}(x) 
               ( \slash{\partial} + m + ie\slash{A}(x) ) 
               \psi^{f}(x)\}. \nonumber
\end{eqnarray}
Bosonization of the model leads in the strong coup\-ling limit 
to the Sine-Gordon model, plus corrections \cite{coleman_76}.

Since the particle spectrum of the Sine-Gordon model is known 
analytically, one can derive that the particle spectrum of
the massive two-flavour Schwinger model 
consists of a pseudo\-scalar isotriplet (``pion'') with G-parity 
G=+1 and a mass $m_\pi$ according to
\begin{equation}
  \frac{m_\pi}{e} = 2.066 \left( \frac{m}{e} \right) ^{2/3},
\label{eq:m_pi(m)}
\end{equation}
and a scalar isosinglet with G=+1 and mass $m_S = \sqrt{3}m_\pi$
\cite{coleman_76}.
In the Sine-Gordon model also the elastic scattering phases have 
been calculated \cite{zamolodchikov_79} and will form the basis for 
our numerical tests. 

{}From the corrections to the Sine-Gordon model only the 
``$\eta$''-particle (a pseudoscalar isosinglet, G=--1) with mass
$m_\eta \approx \sqrt{2} e/ \sqrt{\pi}$ is known \cite{coleman_76}.

\begin{table*}[tb]
\setlength{\tabcolsep}{0.2pc}
\caption{Connection between the irreducible representations 
         of CTS and LTS}
\label{tab:LTS-CTS}
\begin{tabular}{|ccc||ccc|ccc|ccc|ccc|ccc|ccc|}\hline 
  \multicolumn{3}{|c||}{LTS} & \multicolumn{18}{c|}{CTS} \\
  \hline \hline
  \multicolumn{3}{|c||}{} & \multicolumn{6}{c|}{} & 
    \multicolumn{12}{c|}{}\\
  \multicolumn{3}{|c||}{$\Delta^{\sigma_1\sigma_I\sigma_C}$} & 
    \multicolumn{6}{c|}{$\bar{\Delta}_{\bar{D}}^{\sigma_P\sigma_G}$
    (rank 2)} & 
    \multicolumn{12}{c|}{$\bar{\Delta}_{\bar{D}}^{\sigma_P\sigma_G}$
    (rank 4)}\\
  \multicolumn{3}{|c||}{} & \multicolumn{6}{c|}{}  & 
    \multicolumn{12}{c|}{}\\
  $\sigma_1$ & $\sigma_I$ & $\sigma_C$ & $\bar{D}$ & $\sigma_P$ & 
    $\sigma_G$ & $\bar{D}$ & $\sigma_P$ & $\sigma_G$ & $\bar{D}$ & 
    $\sigma_P$ & $\sigma_G$ & $\bar{D}$ & $\sigma_P$ & $\sigma_G$ & 
    $\bar{D}$ & $\sigma_P$ & $\sigma_G$ & $\bar{D}$ & $\sigma_P$ & 
    $\sigma_G$ \\ 
  \hline  
  --&--&--&{ \aa{0.05} }&+ &--&{ \ab{0.05} }&--&--&
    { \aaaa{0.05} }&--&--&{ \aaaa{0.05} }&+ &--&
    { \aaab{0.05} }&--&--&{ \aaab{0.05} }&+ &-- \\
  --&--&+ &{ \aa{0.05} }&+ &+ &{ \ab{0.05} }&--&+ &
    { \aaaa{0.05} }&--&+ &{ \aaaa{0.05} }&+ &+ &
    { \aaab{0.05} }&--&+ &{ \aaab{0.05} }&+ &+ \\
  --&+ &--&{ \aa{0.05} }&--&--&{ \ab{0.05} }&+ &-- &
    { \aaaa{0.05} }&--&--&{ \aaaa{0.05} }&+ &--&
    { \aaab{0.05} }&--&--&{ \aaab{0.05} }&+ &-- \\
  --&+ &+ &{ \aa{0.05} }&--&+ &{ \ab{0.05} }&+ &+ &
    { \aaaa{0.05} }&--&+ &{ \aaaa{0.05} }&+ &+ &
    { \aaab{0.05} }&--&+ &{ \aaab{0.05} }&+ &+ \\
  + &--&--&{ \aa{0.05} }&--&--&{ \aa{0.05} }&+ &--&
    { \aaaa{0.05} }&--&--&{ \aaaa{0.05} }&+ &--&
    { \aaab{0.05} }&+ &--&{  \aabb{0.05} }&--&-- \\
  + &--&+ &{ \aa{0.05} }&--&+ &{ \aa{0.05} }&+ &+ &
    { \aaaa{0.05} }&--&+ &{ \aaaa{0.05} }&+ &+ &
    { \aaab{0.05} }&+ &+ &{  \aabb{0.05} }&--&+  \\
  + &+ &--&{ \aa{0.05} }&--&--&{ \aa{0.05} }&+ &-- &
    { \aaaa{0.05} }&--&--&{ \aaaa{0.05} }&+ &--&
    { \aaab{0.05} }&--&--&{  \aabb{0.05} }&+ &--  \\
  + &+ &+ &{ \aa{0.05} }&--&+ &{ \aa{0.05} }&+ &+ &
    { \aaaa{0.05} }&--&+ &{ \aaaa{0.05} }&+ &+ &
    { \aaab{0.05} }&--&+ &{  \aabb{0.05} }&+ &+  \\
  \hline
\end{tabular}
\end{table*}

The lattice formulation of the Schwinger model with staggered 
fermions and compact Wilson action for the gauge field is given by:
\begin{eqnarray}
  S_{lat}\hspace{-2mm}&=&\hspace{-2mm}\beta \sum_{P} (1 - Re(U_{P}))+ 
\label{eq:S_lat}\\
          & &\hspace{-2mm}\sum_{x,\mu} \frac{1}{2} \eta_{\mu}(x) 
              \{ \bar{\chi}(x) U_{\mu}(x) \chi(x+e_\mu) - \nonumber\\
          & &\hspace{-2mm}\bar{\chi}(x+e_\mu)U^*_{\mu}(x)\chi(x) \} 
              + m \sum_x \bar{\chi}(x)\chi(x). \nonumber
\end{eqnarray}
Energy eigenstates are classified according to irreducible 
representations of the group of symmetry transformations leaving 
the time slices fixed.
The representation $\bar{\Delta}_{\bar{D}}^{\sigma_P\sigma_G}$
of the continuum time slice group (CTS) is characterized by the 
representation $\bar{D}$ of the SU(2) and the quantum numbers 
$\sigma_P$ and $\sigma_G$ associated with parity and G-parity.
The representation $\Delta^{\sigma_1\sigma_I\sigma_C}$ of the 
lattice time slice group (LTS) is characterized by the quantum 
numbers $\sigma_1$, $\sigma_I$ and $\sigma_C$, which correspond 
to shift in space, inversion and charge conjugation on the lattice. 
Restriction of an irreducible CTS representation to the subgroup LTS
will in general lead to reducible representations. 
For the one-(two-)particle operators, which transform according to 
the representations $\bar{D}$ of rank 2 (4), every lattice symmetry 
sector couples to two (four) continuum sectors, as shown in 
Table~\ref{tab:LTS-CTS}. 

\begin{figure}[t]
\begin{center}
\vspace{-5mm}
\epsfig{file=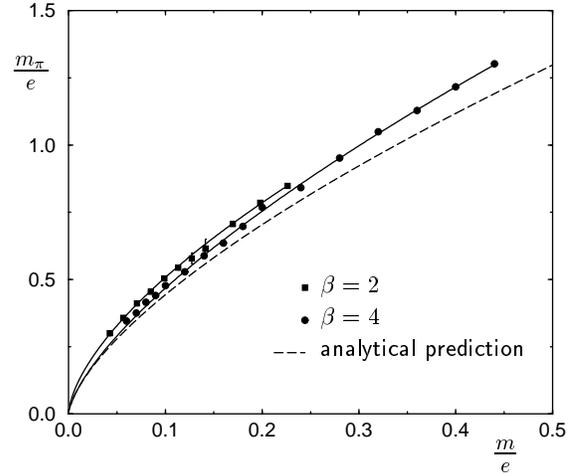,bbllx=61bp,bblly=286bp,bburx=274bp,bbury=475bp}%
\vspace{-5mm}
\caption{The dependence of $m_\pi$ on the bare mass $m$, 
         measured for different values of $\beta$ 
         ($\beta=1/e^2$, $a \equiv 1$) and compared to the 
         analytical prediction (${\rm L}\times {\rm T}=32\times 32$).}
\vspace{-10mm}
\label{fig:m_pi(m)-b2b4}
\end{center}
\end{figure}

\section{FIRST NUMERICAL RESULTS}

We first verified the analytical prediction for the relation of the
pion mass to the bare mass of the fermions (see 
Fig.~\ref{fig:m_pi(m)-b2b4}). 
We have good agreement between the analytical formula and our 
simu\-lations: for $\beta$ = 4 we extract an exponent of 0.689(10) 
as compared to 2/3 of eq.~(\ref{eq:m_pi(m)}).
The deviation between the measured points and the ana\-lytical
curve decreases with increasing $\beta$, which can be explained by 
the fact that the continuum limit corresponds to 
$\beta \rightarrow \infty$. 
This is supported by additional measurements for higher values of 
$\beta$, up to $\beta$ = 10, for which the points tend towards 
the analytically predicted curve.

\begin{figure}[t]
\begin{center}
\vspace{-5mm}
\epsfig{file=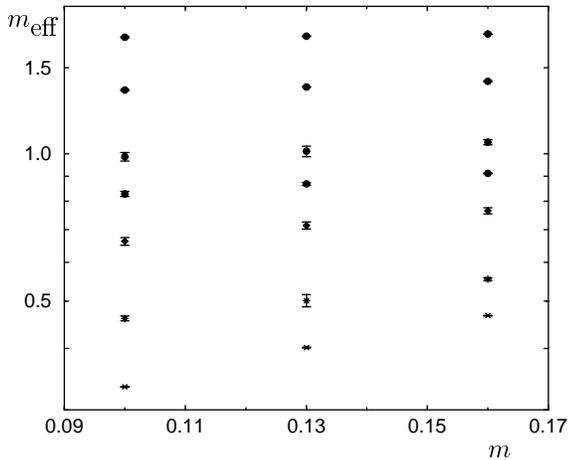,bbllx=61bp,bblly=286bp,bburx=274bp,bbury=475bp}%
\vspace{-5mm}
\caption{The particle spectrum of the Schwinger model for
         $\beta=7$ ($24\times 32$). 
         The lowest state corresponds to the ``pion''.}
\vspace{-10mm}
\label{fig:m_eff(m)-b7}
\end{center}
\end{figure}

In Fig.~\ref{fig:m_eff(m)-b7} we show the particle content 
of the model found by us after the investigation of almost all
lattice symmetry sectors. 
Unfortunately we are not yet able to identify numerically the 
analytically predicted singlet states mentioned above. 

\begin{figure}[t]
\begin{center}
\vspace{-5mm}
\epsfig{file=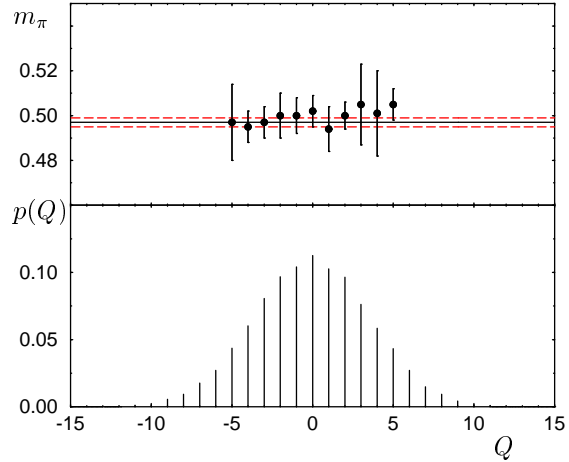,bbllx=61bp,bblly=286bp,bburx=274bp,bbury=475bp}%
\vspace{-5mm}
\caption{Dependence of $m_\pi$ on the topological charge $Q$
         at $\beta=1$ and $m=0.1$ ($16 \times 32$). 
         Measurements on configurations with definite topological 
         charge are compared to the measurement on all configurations 
         (indicated by the lines).}
\vspace{-5mm}
\label{fig:m_pi(Q)+p(Q)-b1m01}
\end{center}
\end{figure}

With increasing $\beta$, a problem related to the topological
properties of the model arises: the tunnelling rate between sectors 
of different topological charges decreases strongly, so that we are
facing an ergodicity problem. To estimate its impact we analyzed 
the dependence of the pion mass on the topological charge 
\begin{equation}
  Q = \frac{1}{2\pi}\sum_P \Theta_P
\label{eq:def-Q}
\end{equation}
of the configurations by comparing measurements on 
configurations associated with different topological charges 
(see Fig.~\ref{fig:m_pi(Q)+p(Q)-b1m01}).
Within the error bars, no significant deviation is visible.

Because of the unexpected complexity of the one-particle spectrum 
for the massive Schwinger model we have not yet been able to determine
scattering phases satisfactorily.

\section{ACKNOWLEDGEMENT}

Helpful discussions with M. G\"ockeler are gratefully acknowledged.
Furthermore we wish to thank the HLRZ J\"ulich and the Rechenzentrum 
of the RWTH Aachen for providing the necessary computer time.

\end{document}